\pdfoutput=1
\documentclass[aps,prx,reprint,twocolumn,groupedaddress,showpacs,svgnames,x11names]{revtex4-1}
\usepackage[utf8]{inputenc}
\usepackage[T1]{fontenc}

\usepackage{color}

\usepackage{braket}
\usepackage{amsmath}
\usepackage{amssymb}
\usepackage{csquotes}
\usepackage{bm}
\usepackage{graphicx}
\usepackage{epstopdf}
\usepackage{subfig}
\usepackage{tikz}
\usetikzlibrary{calc}

\usepackage{xcolor}
\usepackage{hyperref}
\usepackage{bookmark}
\usepackage{array}

\newcommand*{\I}{\mathrm{i}}
\newcommand*{\ttrois}{T$_{3}$ }

\begin{document}

\title{Minimal conductivity, topological Berry winding and duality in three-band semimetals}

\author{Thibaud Louvet}
\author{Pierre Delplace}
\author{Andrei A. Fedorenko}
\author{David Carpentier}

\affiliation{Laboratoire de Physique, École Normale Supérieure de Lyon, 47 allée d'Italie, 69007 Lyon, France}

\date{March 13, 2015}


\maketitle

{\bf
The physics of massless relativistic quantum particles has recently arisen
in the electronic properties of solids following the discovery of graphene.
Around the accidental crossing of two energy bands, the electronic excitations are described by a Weyl equation initially derived for ultra-relativistic
 particles. Similar three and four band semimetals have recently been discovered in two and three dimensions.
Among the remarkable features of graphene are the characterization of the band crossings by a topological Berry winding,
leading to an anomalous quantum Hall effect, and a finite minimal conductivity at the band crossing while the electronic density vanishes.
Here we show that these two properties are intimately related: this result paves the way to a direct measure of the topological nature of a semi-metal.
By considering three band semimetals with a flat band in two dimensions, we find that
only few of them support a topological Berry phase. The same semimetals are the only ones displaying a non vanishing
minimal conductivity at the band crossing. The existence of both a minimal conductivity and a topological robustness  originates from properties
of the underlying lattice, which are
encoded not by a symmetry of their Bloch Hamiltonian, but by a  duality.
}

Electronic excitations which satisfy ultra relativistic quantum physics emerge around the band crossing of semimetals. While graphene constitutes
 the canonical example of such a phase, the discovery of topological insulators opened the route to other realizations~\cite{Murakami:2007}:
 such semimetals exist at the transitions between topological and trivial insulators. Theoretical proposals were initiated by the identification of
  crystalline symmetries stabilizing these critical phases both in two and three dimensions~\cite{Young:2012}. This idea
  turned out to be particularly successful: recently,
a Weyl semimetal corresponding to two band crossings in three spatial dimensions was discovered in TaAs~\cite{Lv:2015,Xu:2015}.
The existence of a three band crossing  semimetal was proposed in a two dimensional carbon allotrope
 SG-10b  \cite{Wang:2013b} and square MoS$_2$ sheet~\cite{Li:2014}, and discovered in three dimensions in HgCdTe~\cite{Orlita:2014}.
 Stable four band crossing semimetals in three dimensions, denoted Dirac phases,  were proposed theoretically and experimentally identified in
 Na$_3$Bi~\cite{Wang:2012,Liu:2014,Xu:2015} and in Cd$_3$As$_2$~\cite{Wang:2013,Neupane:2014,Borisenko:2014}, 
 and predicted  in $\beta$-cristobalite BiO$_2$~\cite{Young:2012} and distorted spinels~\cite{Steinberg:2014}.

In the case of graphene, the relativistic nature of the electronic excitations translates into remarkable transport properties including
an anomalous half-integer Hall effect with half integer plateaus and a non vanishing minimal conductivity exactly at the band crossing
\cite{Novoselov:2005,Zhang:2005,Twordzylo:2006,Katsnelson:2006b}.
 The anomalous  half-integer Hall effect is related to the topological properties of the band crossing: when winding around the crossing point, an
 electron picks up a quantized so-called Berry phase. This Berry phase is at the origin of the half-integer nature of the quantum Hall effect
 in graphene \cite{Zhang:2005,Novoselov:2006}, but also characterizes the topological property of the semi-metallic phase encoding its
 robustness towards gap opening perturbations \cite{Murakami:2007}. On the other hand the minimal conductivity at the band crossing
 was associated to the Zitterbewegung of Dirac particles, an intrinsic agitation characteristic of ultra relativistic particles which leads
 to diffusive motion even in perfectly clean samples \cite{Twordzylo:2006,Katsnelson:2006}.

 In this article we consider these characteristic signatures of semimetals beyond the two-band crossing situation of graphene.
 When a third locally flat band is present at the crossing, none of these properties is necessarily enforced. Indeed, we find that 
 the exact same models  possess both a non vanishing minimal conductivity and a topological Berry winding at the crossing. Hence these
 properties are not hallmarks of a relativistic energy spectrum: they encode phase  properties of the electronic wave functions. In the case
 of graphene, this phase winding originates from the hexagonal lattice of carbon atoms. We will show that the lattice properties at
 the origin of these remarkable transport signatures are encoded not by a standard symmetry constraints of Bloch Hamiltonians, but by a duality
 transformation. We identify two duality classes, corresponding to the two  signatures: the existence or not of both a non-vanishing minimal conductivity and
a topological Berry winding.

\medskip
{\bf Chirality or sublattice symmetry}.
\begin{figure}[h!]
\includegraphics[width=6cm,angle=-90]{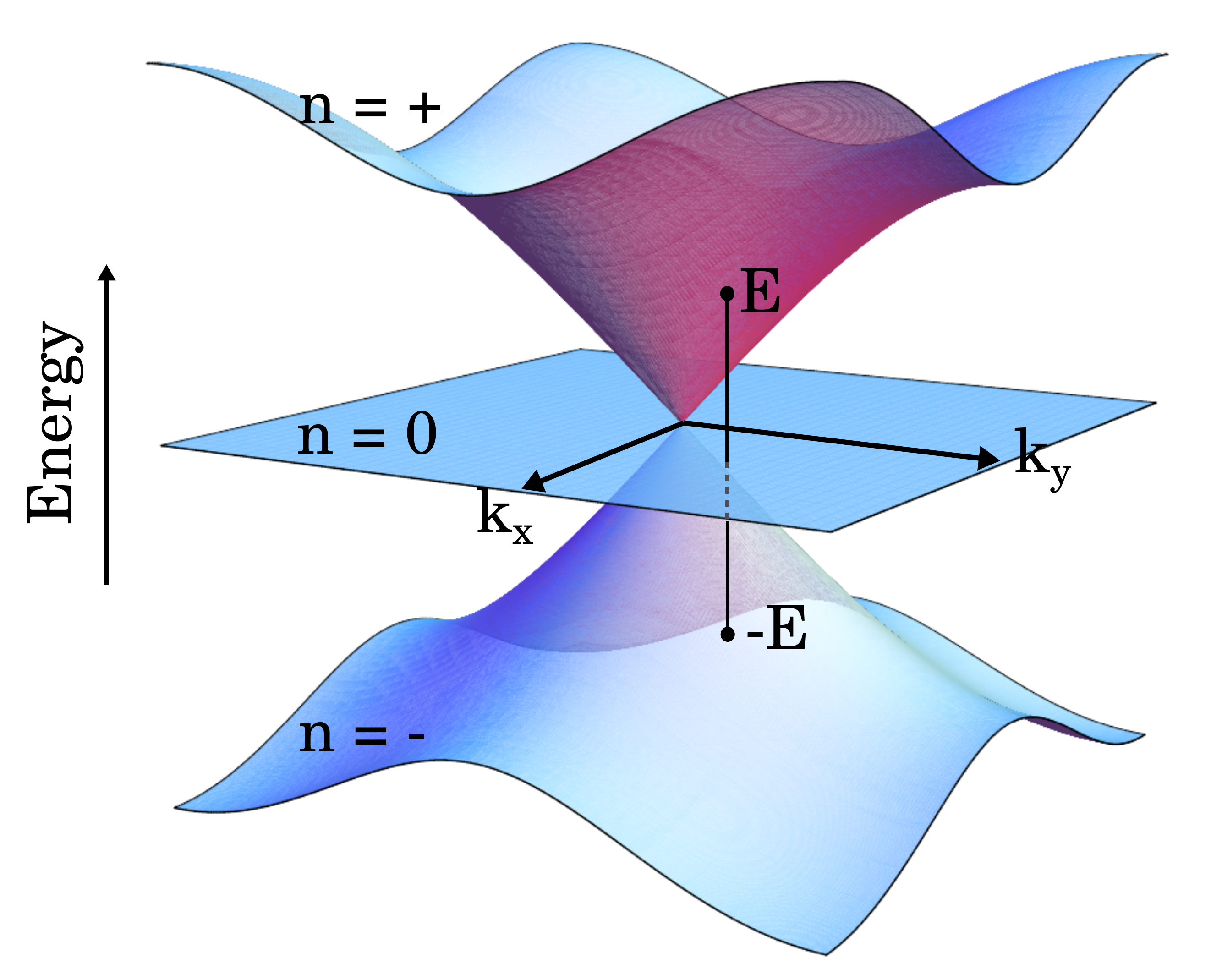}
\caption{Energy spectrum of a three-band chiral semimetal
which shows a symmetry $E(\vec{k})\rightarrow -E(\vec{k})$. Such a spectrum
consists of two linear energy bands $n=\pm$ and a flat band $n=0$.}
\label{fig:cone}
\end{figure}
\begin{figure*}[ht!]
\includegraphics[width=10cm,angle=-90]{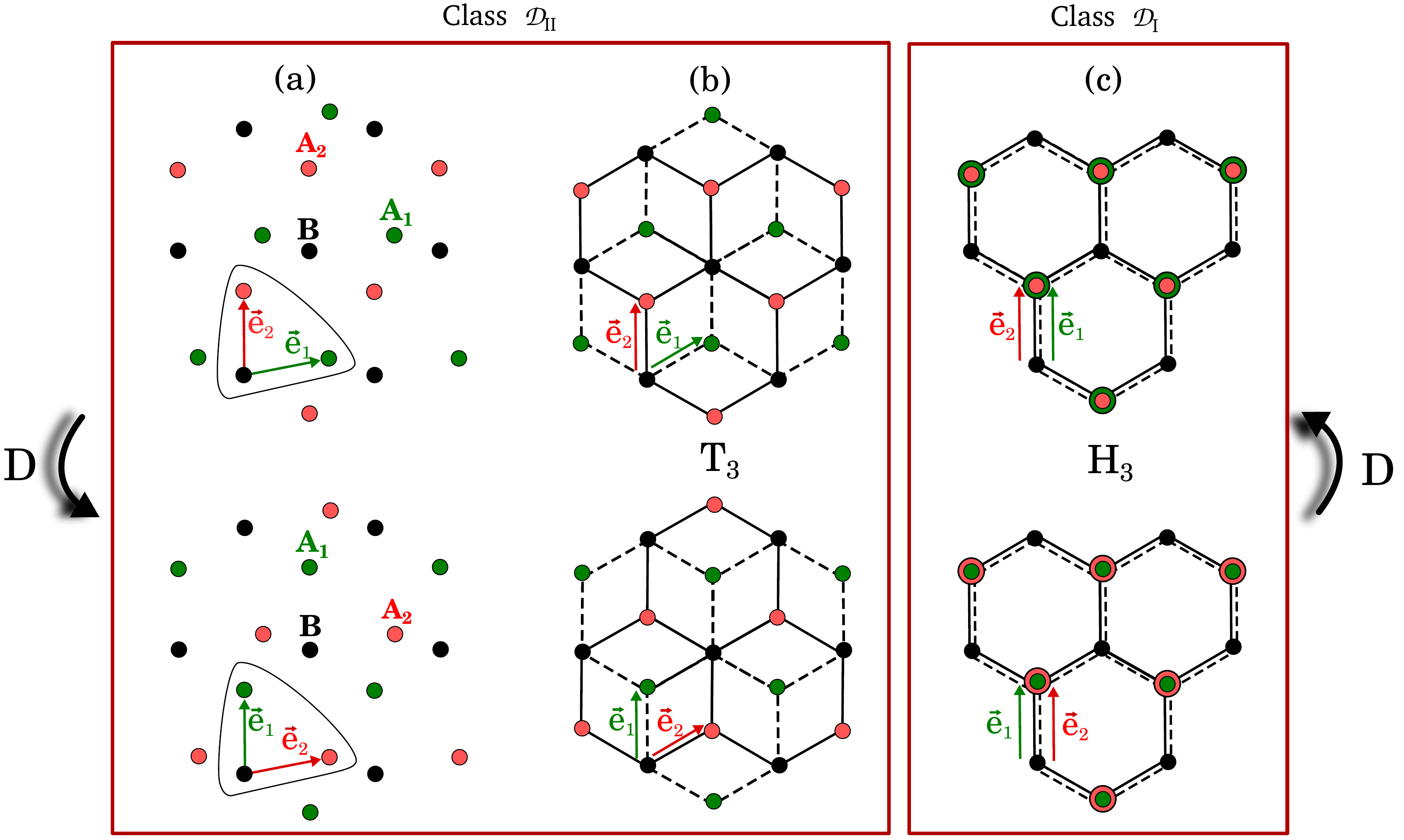}
\caption{Triangular Bravais lattices with three orbitals ($A_1$, $A_2$ and $B$) per unit cell.
The position of the $A_1$ orbitals is chosen arbitrary (a), in the center of the $BA_1$ hexagons (b), or at the same location than the $A_2$ orbitals (c)
The duality transformation $\mathcal{D}$ exchanges the location of the orbitals $A_1$ and $A_2$ and the hopping amplitudes symbolized
by full/dashed lines.
The original T$_3$ lattice is recovered after an inversion in (b) whereas the dual and the original lattices coincide for the H$_3$ lattice in (c).}
\label{fig:lattices}
\end{figure*}
We consider three-band semimetals in two dimensions characterized by  two linearly crossing energy bands $n=\pm $ and a third locally flat band
 $n=0$, represented  in Fig.~\ref{fig:cone}.
 This corresponds to the situation of HgCdTe  in three dimensions~\cite{Orlita:2014}, or the carbon allotrope SG-10b~\cite{Wang:2013b}
 and square MoS$_2$ sheet~\cite{Li:2014} in two dimensions.
 The energy spectrum $E(\vec{k})$ as a function of the momentum $\vec{k}$ exhibits the symmetry $E(\vec{k})\rightarrow -E(\vec{k})$
 at least locally around the crossing point.
This spectrum symmetry naturally originates from a chiral symmetry of the corresponding (low energy) Hamiltonian. This symmetry
 is represented by a unitary operator $C$ that \emph{anticommutes} with the Hamiltonian: $\mathcal{H}=-C\mathcal{H} C$.

 An explicit chiral operator can be defined when considering the pedagogical examples of tight-binding Hamiltonians defined on lattices.
 Similarly to graphene, only nearest neighbor couplings can be kept when focusing around the band crossing points.
 In this case chiral symmetry corresponds to a sub-lattice symmetry:  couplings are only present between the two sub-lattices $A$ and $B$ of
  a bipartite lattice.  
  This is the case of the nearest neighbor description of graphene on the honeycomb lattice. In the case we consider in this article,
  three bands crossing implies the existence of three orbitals distributed on three Bravais lattices  $A_1, A_2$ and $B$ of same geometry,
  as shown on Fig.~\ref{fig:lattices}. Chiral symmetry is satisfied if  the only couplings $t_1, t_2$ relevant at low energy 
  are between orbitals on the $B$ and the  $A_1, A_2$ lattices whereas $A_1$ and $A_2$ stay uncoupled.
  The corresponding Bloch Hamiltonian  in the orbital basis $(A_1,A_2,B)$ is written
\begin{equation}
 H(t_1,t_2;\vec{k})=\left( \begin{array}{ccc}
             0 & 0 & t_1f_1 (\vec{k}) \\
             0 & 0 & t_2f_2 (\vec{k}) \\
             t_1f_1^* (\vec{k}) & t_2f_2^* (\vec{k}) & 0
            \end{array} \right)\ .
\label{eq:general hamiltonian}
\end{equation}
 Such a Hamiltonian anti-commutes 
 with a chirality operator $C = \textrm{diag}(1,1,-1)$.
The complex functions $f_j(\vec{k})=|f_j(\vec{k})|e^{i\phi_j(\vec{k})}$  encode the geometry of the  lattice of couplings.
Their amplitudes determine the spectrum of the semimetal: $E_0(\vec{k})=0, E_{\pm}(\vec{k})=\pm 
(t_1^2|f_1(\vec{k})|^2+t_2^2|f_2(\vec{k})|^2 )^{\frac12}$. A three band crossing occurs when $f_1$ and $f_2$ vanish simultaneously at a point $\vec{K}$ in the Brillouin zone.
In this article, we focus on properties which depend on the phases $\phi_j(\vec{k})$ and which are thus \emph{independent} of the spectrum provided
a band crossing occurs.

Quite generally, we will consider chiral symmetric Bloch Hamiltonians describing a  three-band semimetal with band crossing at point  $\vec{K}$,
and whose linear expansion around the crossing takes the form
\begin{multline}
 H(\vec{K}+{\vec q})= \\
 	\left( \begin{array}{ccc}
             0 & 0 & \Lambda_{11}q_x + \Lambda_{12}q_y\\
             0 & 0 & \Lambda_{21}q_x + \Lambda_{22}q_y \\
             \Lambda_{11}^*q_x + \Lambda_{12}^*q_y & \Lambda_{21}^*q_x + \Lambda_{22}^*q_y & 0
	\end{array} \right) \ .
\label{eq:general linear hamiltonian}
\end{multline}
Such a Hamiltonian is entirely parametrized by a matrix $\Lambda=\{ \Lambda_{ij}\}$  of complex coefficients.
 The phase of the coefficients $\Lambda_{ij}$ encodes the geometry of the underlying lattice. The corresponding geometrical
 constraint on these phases must be independent of the amplitude of couplings between the orbitals. Hence it cannot result in a
 symmetry of the Hamiltonian: we show that it corresponds to a duality in a manner analogous to the Kramers-Wannier duality of statistical
 mechanics models which relates models with different coupling amplitudes on different lattices~\cite{Kramers:1941}.

 \medskip
 {\bf Lattice geometry and duality constraints}.
 The geometry of
 the lattice can be described by two vectors  $\vec{e}_1$ and $\vec{e}_2$ relating a vector of the lattice $B$ to neighbor sites of
 the  $A_1$ and $A_2$ lattices, as shown on figure \ref{fig:lattices}.
 The duality transformation $\mathcal{D}$ exchanges the lattices $A_1$ and $A_2$, or equivalently the vectors $\vec{e}_1$ and $\vec{e}_2$,
  while simultaneously exchanging the couplings between $B$ and $A_1$ orbitals with couplings between  $B$ and $A_2$ orbitals.
  Quite generally,  this duality which is an involutive transformation {\it i.e.} ${\cal D}^2=\mathbf{I}$,  
 relates a Hamiltonian $\mathcal{H}$ on a lattice $\mathcal{L}$ to a Hamiltonian $\tilde{\mathcal{H}}$
 on a \emph{different} lattice $\tilde{\mathcal{L}}$. However, on symmetric lattices where initial and dual lattices
 $\mathcal{L},\tilde{\mathcal{L}}$ are related by a geometrical transformation $\mathcal{R}$, this duality translates into constraints on
 Hamiltonians defined on the same lattice (or same Hilbert space). In this case, and focusing for simplicity on nearest neighbor Hamiltonians,
the duality transformation can be recast into the form
\begin{equation}
(DU) H (t_2,t_1; \mathcal{R}\vec{k}) (DU)^{-1} = H (t_1,t_2;\vec{k})
\label{eq:geometric duality}
\end{equation}
where $U$ is a unitary operator, $\mathcal{R}$ is the symmetry relating initial and dual lattices and $D$ 
the operator swapping $A_1$ and $A_2$ orbitals:
\begin{equation}
D=\left( \begin{array}{ccc}
                                             0 & 1 & 0 \\
                                             1 & 0 & 0 \\
                                             0 & 0 & 1
                                            \end{array} \right) \ .
 \label{eq:dualityop}
\end{equation}
A very special case, which we call the duality class $\mathcal{D}_{\textrm{I}}$ corresponds to the situation where two orbitals lie on the same
geometrical lattice, {\it i.e.} when $\vec{e}_1 =\vec{e}_2\neq \vec{0}$.
In this class, the duality transformation simplifies and $U$ and $\mathcal{R}$
reduce to the identity. In this case, Bloch Hamiltonians encode the  geometrical properties of a bipartite lattice, whereas in the
other case, which we denote the  duality class $\mathcal{D}_{\textrm{II}}$, the underlying lattices are either Bravais lattices or possess three
distinct sublattices.
 This duality restricts 
 the form of the chiral tight-binding Hamiltonian  (\ref{eq:general hamiltonian}): in the class $\mathcal{D}_{\textrm{I}}$
 we have  $f_1 (\vec{k}) = f_2 (\vec{k})$ while a much weaker constrain  $f_1 (\vec{k}) = f_2 (\mathcal{R}\vec{k})$ holds 
  in class $\mathcal{D}_{\textrm{II}}$ for symmetric lattices.
Generalizing this constrain to a local Bloch Hamiltonian (\ref{eq:general linear hamiltonian}) around a three band crossing, the
duality in class $\mathcal{D}_{\textrm{I}}$ implies the condition
\begin{equation}
 \frac{\Lambda_{21}}{\Lambda_{11}} = \frac{\Lambda_{22}}{\Lambda_{12}} \equiv \lambda \ ,
\label{eq:lambda}
\end{equation}
while generically it only relates Hamiltonian at different crossing points in class $\mathcal{D}_{\textrm{II}}$.  
   In the following we show that Hamiltonians belonging to the duality
  class $\mathcal{D}_{\textrm{I}}$ describe the only three-band semimetals possessing quantized topological Berry windings which are also those whose
  conductivity does not vanish at the band crossing and display a pseudo-diffusive regime.

  To illustrate this relation, let us discuss two nearest-neighbor lattice models of semi-metals belonging to both classes.
A natural model in class $\mathcal{D}_{\textrm{I}}$, inspired by graphene, corresponds to a honeycomb lattice with two orbitals on one
of the two sub-lattices, shown on Fig.~\ref{fig:lattices}~(c).
The three bands of this model, which we call H$_3$, cross at points $\vec{K}$ and $\vec{K}'$ of the Brillouin zone. Around those points,
the Bloch Hamiltonian takes the form~\eqref{eq:general linear hamiltonian} with a matrix of coefficients
\begin{equation}
 \Lambda_{\textrm{H}_3}= \frac{3a}{2} \left(\begin{array}{cc}
           t_1 & -\I t_1 \\
           t_2 & -\I t_2
          \end{array} \right),
 \label{eq:lambda-3/2}
\end{equation}
with $a$ being the honeycomb lattice spacing, while the characteristic energy scale of nearest neighbor couplings is encoded into $t=\sqrt{t_1^2+t_2^2}$.
 This model satisfies the condition \eqref{eq:lambda} with $\lambda = -\I$ and belongs to the duality class $\mathcal{D}_{\textrm{I}}$.
The \ttrois model \cite{Raoux:2014} is again defined on a honeycomb lattice but with additional orbitals $A_2$ at the center of each hexagon as
shown on Fig.~\ref{fig:lattices} (b).
These orbitals are coupled by an amplitude $t_2$  only to the $B$ sub-lattice of the honeycomb lattice, while
$A_1$ and $B$ orbitals on the honeycomb lattices are coupled by $t_1$.
This model belongs to the duality class $\mathcal{D}_{\textrm{II}}$, with the
 inversion $\mathcal{R} \vec k = - \vec k$ relating initial and dual lattices.
 Indeed, the Bloch Hamiltonian linearized around the band touching point $\vec K$ is written in the form~\eqref{eq:general linear hamiltonian}
with
\begin{equation}
 \Lambda_{\textrm{\ttrois}}= \frac{3a}{2}
 \left(\begin{array}{cc}
                   t_1 & -\I t_1  \\
		 	t_2 & \I t_2  \\
                 \end{array} \right)~,
\label{eq:lambda-t3}
\end{equation}
which does not fulfill the condition \eqref{eq:lambda}. Note that when $t_1=t_2$, this linearized Hamiltonian can be written
in the form $H_{\bf K} ({\bf q}) = \hbar v_F \bf S \cdot  q $, where $S_x$, $S_y$
and $S_z \equiv \text{diag} (1,-1,0)$ satisfy the spin-$1$
algebra $[S_i,S_j]= \I \epsilon_{ijk}S_k$.
Hence the \ttrois model realizes a continuous deformation of spin-$1$ massless fermions, which all belong to the
duality class $\mathcal{D}_{\textrm{II}}$.

We now characterize both the topological properties of the band crossing, as well as the electronic transport properties around the crossing,
 which turn out to be associated to the duality class of the semi-metal.

\medskip
{\bf Topological property of a band crossing}.
\begin{figure*}[!thbp]
 \includegraphics[width=0.9\textwidth]{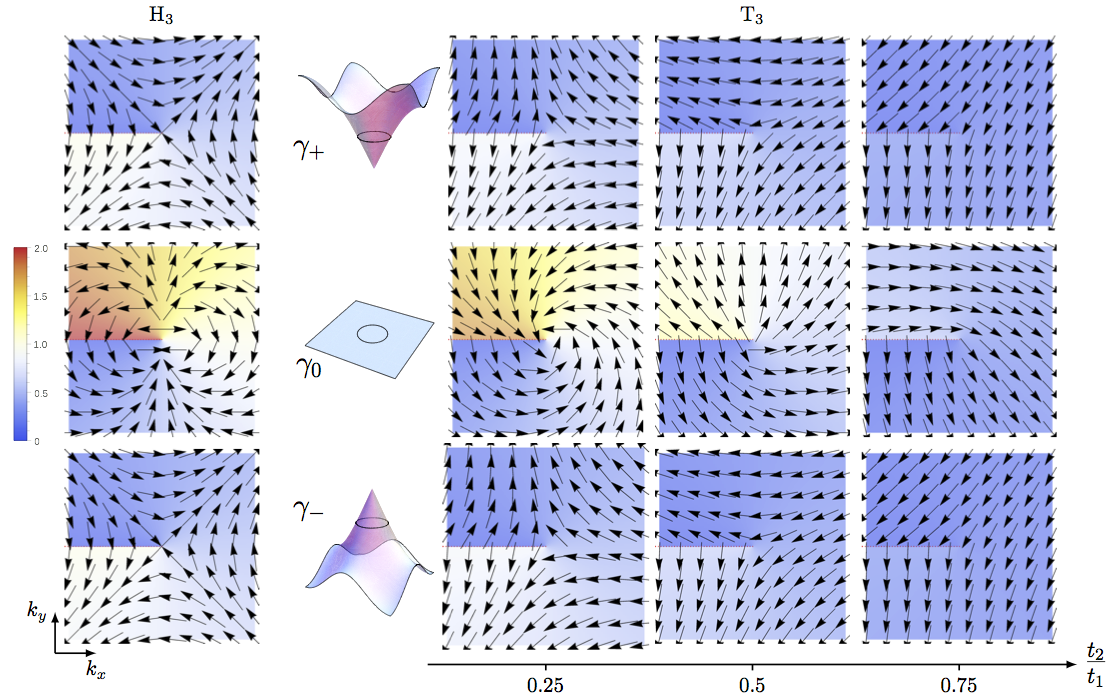}
 \caption{Momentum dependance of the phase associated with the Berry connection for the three energy levels $n=-,0,+$ for nearest neighbors models on the
  H$_3$ and on the \ttrois model for three different values of the ratio of couplings between sub-lattices. A vortex is associated with
 a $\pi$ Berry phase. For the H$_3$ lattice model,  the Berry phase $\gamma_n$ \eqref{eq:gamman} along any closed loop
 winding around the origin is quantized (in units of $\pi$) and signals a topological property of the band crossing points. This is not the case for the
 \ttrois lattice model, where this Berry phase continuously decreases and vanishes as a function of the couplings. }
 \label{fig:winding}
\end{figure*}
The topological properties of band crossings in two dimensional crystals can be described by 
 a topological invariant associated to each band around the crossing point.
 These invariants are the Berry phases $\gamma_n$ (see eq.\eqref{eq:gamman} of Methods)
  acquired by Bloch eigenstates $\Psi_n$ upon winding around the crossing point $\vec{K}$.
By definition this Berry winding is independent of the path winding around the crossing point: it is an homotopic invariant of a given Hamiltonian.
It describes a \emph{topological} property of the semimetal when it is robust against perturbations of the Hamiltonian which do not
lift the band crossing. Such a robustness occurs when this Berry winding is quantized. This is indeed the case in graphene where
$\gamma_{\pm} = \pm 1$.
 For the three-bands semi-metals, these Berry windings can be readily obtained from the diagonalisation of the Hamiltonian 
 \eqref{eq:general linear hamiltonian}. We find that these windings are topological only when the condition~\eqref{eq:lambda} is fulfilled:
 the only
 semimetals characterized by  a topological Berry winding are those of the duality class $\mathcal{D}_{\textrm{I}}$, with values
\begin{equation}
 \gamma_{+} =\gamma_- = \text{sgn}\, \text{Im}\, \lambda \quad , \quad  \gamma_0 = -2\, \text{sgn}\, \text{Im}\, \lambda \ .
\end{equation}
These quantized values of the Berry windings are stable with respect to any perturbation compatible with the duality constraint,
{\it i.e.} which does not break the geometry of the underlying lattice.
Eigenstates of the H$_3$ model \eqref{eq:lambda-3/2} are characterized by Berry phases $\gamma_{\pm}=-1$, $\gamma_0=2$  
as shown in Fig.~\ref{fig:winding}. These windings are in particular robust to variations of hopping amplitudes $t_1,t_2$.
For any chiral symmetric semimetal which does not belong to the $\mathcal{D}_{\textrm{I}}$ class, the Berry phase can take any real value and depends
continuously on deformations of the Hamiltonian. 
This result is illustrated for the \ttrois model \eqref{eq:lambda-t3} in Fig.~\ref{fig:winding}: the Berry phases $\gamma_n$ are shown to vary continuously upon variation of the ratio $t_2/t_1$ of nearest neighbor couplings.

\begin{figure}[!t]
%
	\includegraphics[width=8cm]{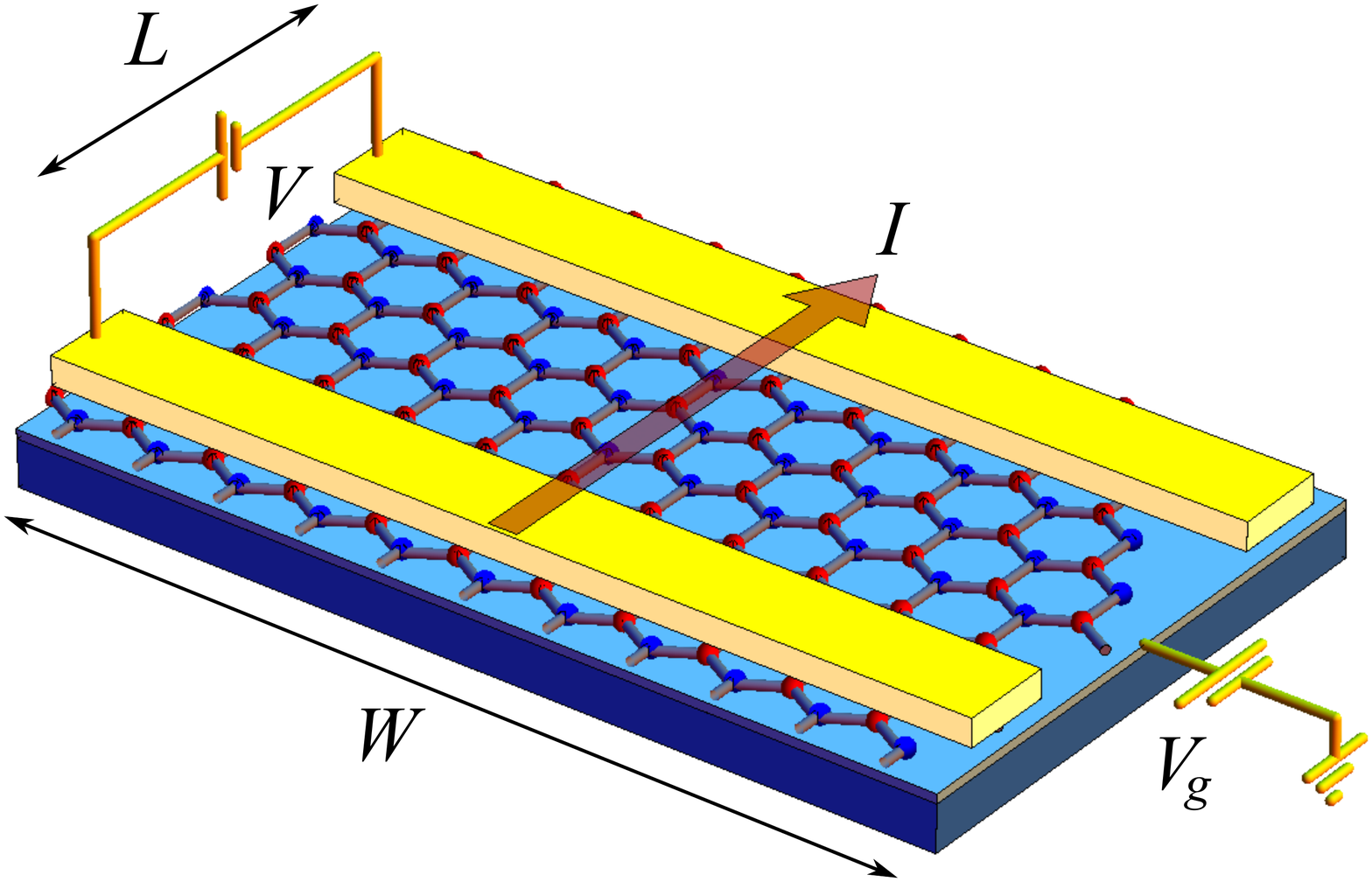}
\caption{ The setup for two-terminal transport measurements:
 a 2D sample of size $W \times L$ with leads attached on opposite sides.
\label{fig:cond3-1} }
\end{figure}
\begin{figure*}[!t]
\subfloat[]{%
	\includegraphics[width=8.5cm]{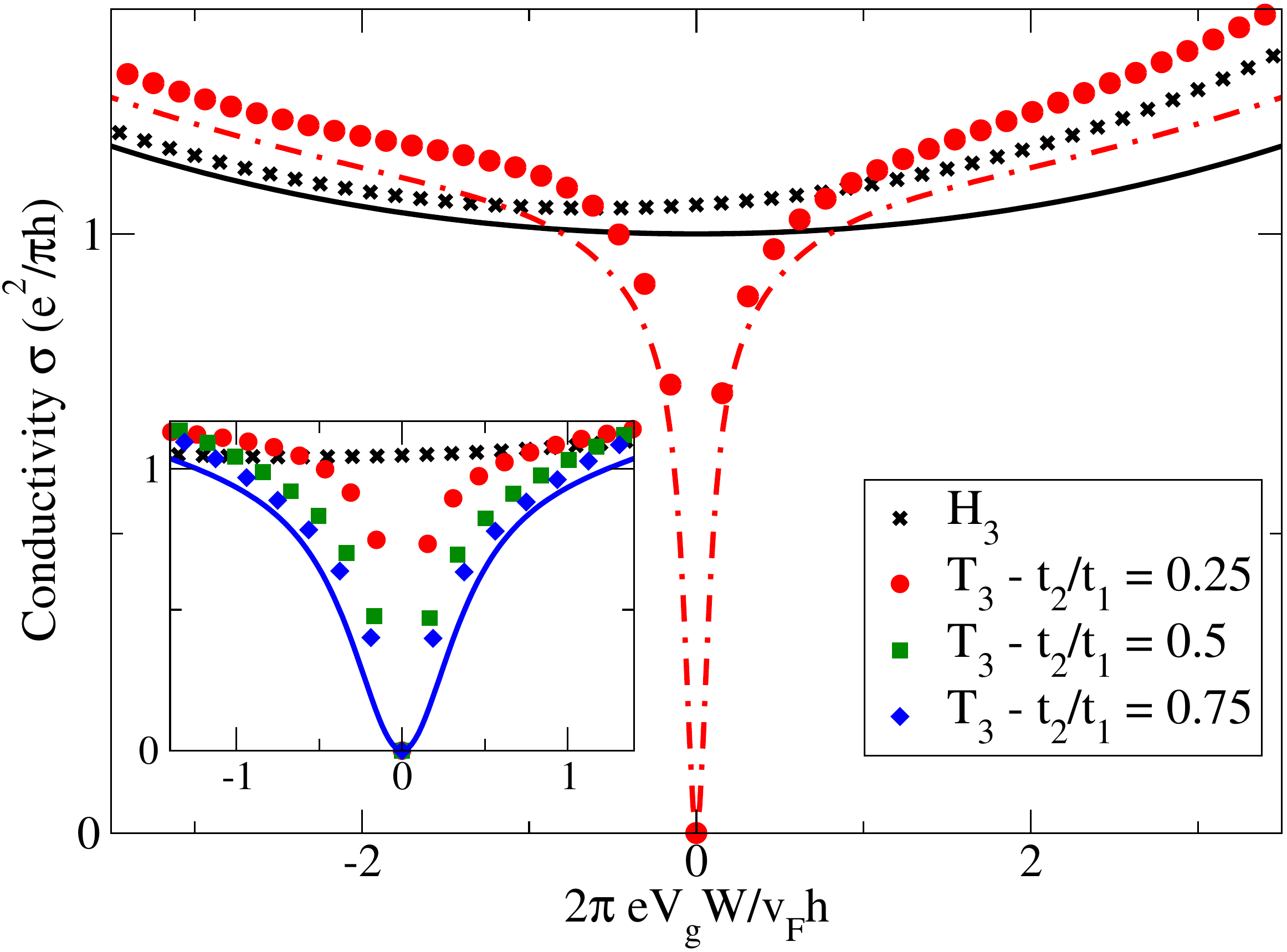}
	\label{fig:cond3-2}}
\subfloat[]{%
    	\includegraphics[width=8.5cm]{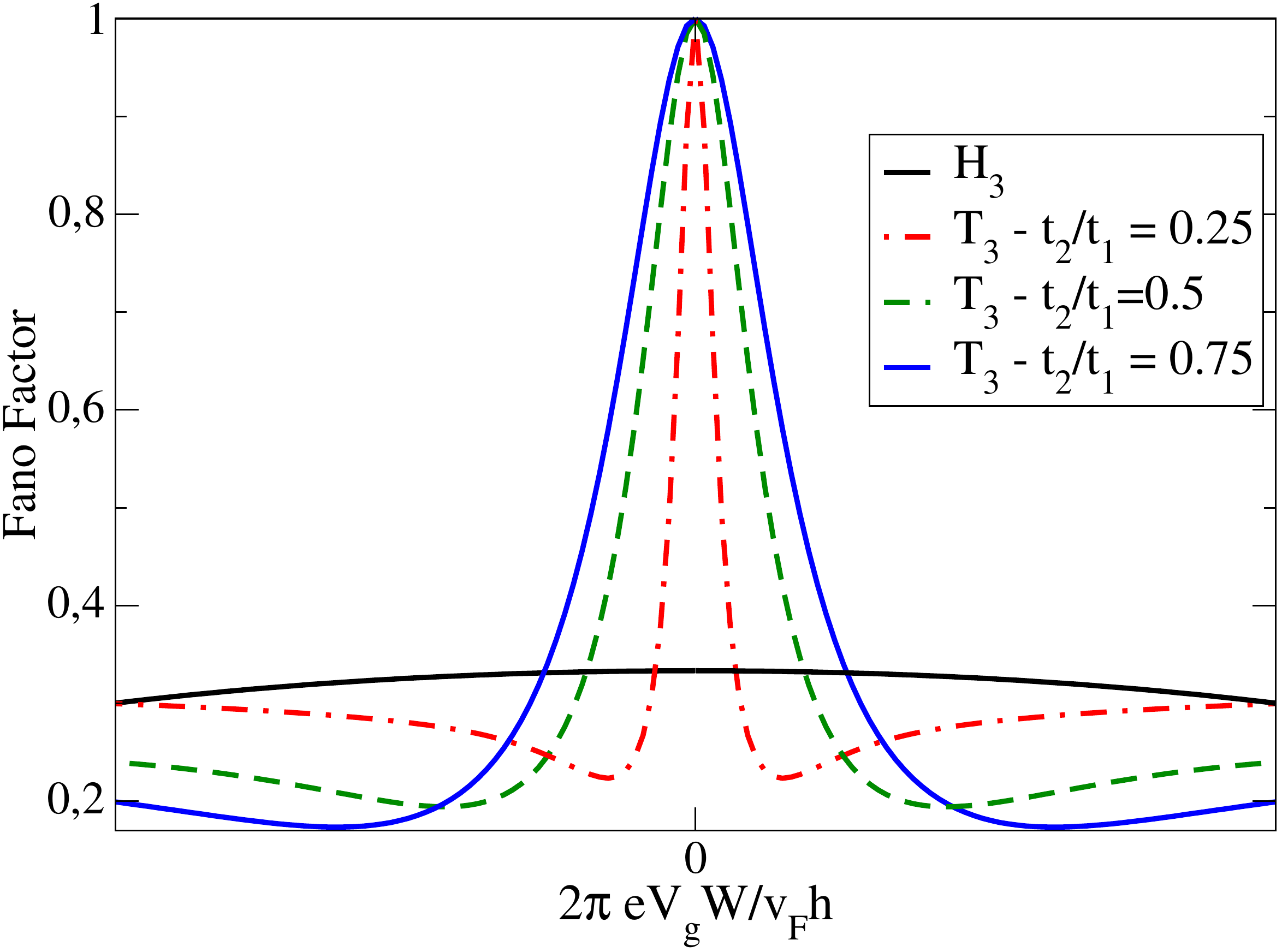}
   	\label{fig:cond3-3}}
\caption{
   Conductivity (a) and Fano factor (b) as a function of a gate potential $V_g$ applied to  the sample of size $L=100, W=300$ in units of lattice spacing $a$. 
  Fermi velocity is given by $v_F=3at/(2\hbar)$. Results for nearest neighbor lattices
  models on the  H$_3$ and \ttrois lattices are shown for various ratio $t_2/t_1$ of couplings between the different lattices. }
\end{figure*}

\medskip
{\bf Minimal conductivity at the band crossing}.
Transport measurements constitute a powerful tool to probe the physical properties in the vicinity of the Fermi energy. We will show that close
to the band crossing electronic transport is related to the phase content of the Hamiltonian (the phases $\phi_{i}(\vec{k}$) of the amplitudes in  
\eqref{eq:general hamiltonian})  and not to the spectrum.
Remarkably, in graphene,
when the Fermi level coincides with the twofold band crossing point, the conductivity of a clean sample was predicted to remain finite
despite a vanishing density of states. This result was first derived by considering the conductivity of a narrow strip of
graphene between two contact electrodes as shown on  Fig.~\ref{fig:cond3-1}.
Let us consider an analogous setup for a three band chiral semimetal, {\it i.e.} a finite sample of length $L$ and width $W$.
Confinement of the sample between the leads gives rise to  zero-energy evanescent states. At the band crossing,
 the conductivity depends entirely on the nature of these evanescent states.
Prior to an explicit calculation of the conductivity (see Methods), it is instructive to consider the current operator
$j_x (\vec k) = \langle \partial_{k_x} H(\vec k) \rangle_\psi$ defined from the tight-binding Bloch Hamiltonian~\eqref{eq:general hamiltonian}.
Introducing the amplitudes $( \psi_{A_1}, \psi_{A_2}, \psi_B)$ of the electronic wavefunction in the three sub-lattices the longitudinal current
can be expressed as
$  j_x (\vec k) = 2~ \text{Re}\, [ \psi_B (\vec k) (\partial_{k_x} f_i (\vec k)
  \psi_{A_i}^* (\vec k) )] $
 and is found to be proportional to the amplitude $\psi_B$ on the $B$ sub-lattice \cite{Hausler:2015}.
This hints that electronic transport at the band crossing will occur provided 
the zero-energy evanescent modes have a non-vanishing component on the $B$ sub-lattice.

As expected from the previous qualitative argument, the existence of a finite minimal conductivity at the threefold
band crossing point is uniquely determined by the non vanishing weight of the wave function on the
hub lattice as we have checked using a Landauer description of transport (see Methods). 
 From \eqref{eq:general linear hamiltonian}, this component is found to satisfy
\begin{equation}
\Lambda .\left( \begin{array}{c}
				      q_x \\
				      q_y
				      \end{array} \right) \psi_B = 0 \ .
\label{eq:psiB}
\end{equation}
Hence the condition of existence of a non-vanishing minimal conductivity at the band crossing is simply 
$ \text{det}\,\Lambda = 0 $.
This constraint exactly coincides with the duality constraint \eqref{eq:lambda} defining the class $\mathcal{D}_{\textrm{I}}$: the only
three band semi-metals with a non-vanishing minimal conductivity are exactly those belonging to this duality class $\mathcal{D}_{\textrm{I}}$.
Moreover for the H$_{3}$ lattice model, this minimal conductivity corresponds exactly to the value
$ \sigma^{\textrm{(min)}} = e^2/(\pi h)$ predicted for graphene. This result remains valid for any model in the dual class $\mathcal{D}_{\textrm{I}}$ with
an isotropic dispersion relation.
 For more general three band semi-metals with an anisotropic dispersion relation, the minimal conductivity 
 depends on the angle 
 between the leads and the principal axes of the dispersion relation. However, the determinant of the corresponding conductivity tensor remains 
 constant given by its isotropic value $\det \overline{\sigma} = (e^2/(\pi h))^2$. 
In contrast, any three-band chiral symmetric semimetal that does not belong
to the $\mathcal{D}_{\textrm{I}}$ class possesses a vanishing conductivity $\sigma^{\textrm{(min)}} = 0$ in every directions.
Beyond the minimal conductivity, the fluctuations of this conductivity can also be considered : their amplitude is encoded in the ratio between the shot noise
power and the averaged current, the so-called
 Fano factor. This factor $F$ takes a constant value $F=1/3$ within the duality class $\mathcal{D}_{\textrm{I}}$, a value already encountered in graphene 
 \cite{Twordzylo:2006}
 and characteristic of diffusive metals  \cite{Beenakker:1997}. Such a result demonstrates that
 for all semi-metals in this class
 the transport through narrow perfectly clean junctions displays the characteristic features of diffusive metals. 

 We have evaluated the conductivity of different lattice models in the geometry of Fig.~\ref{fig:cond3-1}.
 We compare the analytical results to a numerical Landauer approach (see Methods) to check for possible inter-crossing point effects, neglected in the
 analytical approach. A perfect agreement is found between both approaches.
  The results for the H$_3$ model are shown as a  function of the Fermi-energy, or gate potential   $V_g$ on  Fig.~\ref{fig:cond3-2}:
  the conductivity exhibits a plateau around the band crossing point $V_g=0$, corresponding to $\sigma = e^2/ (\pi h)$.
 The results of a similar study for the \ttrois model are also shown on Fig.~\ref{fig:cond3-2} and display a collapse of the
 conductivity  around the band crossing point $V_g=0$ for three different values of the couplings between orbitals (Inset).
 Fig.~\ref{fig:cond3-3} displays the analytical results for the dependance of the Fano factor on the gate potential $V_g$. 
 We show that  for the \ttrois model $F(V_g=0)=1$, whereas for the H$_3$ model the Fano factor 
 reaches the value $F=\frac13$ at the band crossing,   characteristic of a  disordered metal as expected for class $\mathcal{D}_{\textrm{I}}$. 
Finally, let us mention that for graphene the result of the Landauer formula for a narrow junction can be recovered for a long junction 
by an approach based on the Kubo formula~\cite{Ryu:2007}. 
While this equivalence 
remains valid for
three band semimetals in the dual class $\mathcal{D}_{\textrm{I}}$, it does not hold beyond it: we found that the Kubo conductivity for the \ttrois
model diverges at the band crossing, as  opposed to the vanishing
 Landauer conductivity for a narrow junction, in agreement with a previous result in the disordered limit~\cite{Vigh:2013}.

As follows from our results, the occurrence of a transport regime with a non vanishing conductivity at the crossing is not a generic property of linear dispersion 
relations near this crossing, nor a hallmark of relativistic physics of the associated electronic excitations such as the Zitterbewegung.
It is indeed intimately related to the existence of a topological robustness
of this band crossing, originating from lattice-properties encoded into the class $\mathcal{D}_{\textrm{I}}$ condition.
   This result opens the route to a direct probe of topological properties of semimetals through transport measurements around the band crossing.
 In graphene, a quantitative measurement of the conductivity exactly at the band crossing is hampered by fluctuations of the chemical potentials
 induced by a back gate.
 However the recent discovery of three dimensional semimetals changes the perspective: most of these new semi-metallic materials are stoichiometric and
 the Fermi-energy is expected to reside at the band crossing. In that situation the transport will be entirely dominated by the physics at the band crossing and
 will directly probe the topological properties associated with the band crossing points.

Let us put in perspective  the  association between 
minimal conductivity and topological properties raised by our results on three band semimetals in two dimensions.
In two dimensions, we can extend our analysis for a band crossing described by the simple Hamiltonian $H(\vec{q})=S_x q_x + S_y q_y$, with $S_x,S_y$ satisfying a 
spin-$S$ algebra. Again, we find that the Berry topological winding \cite{Dora:2011} and the minimal conductivity are correlated. 
In particular for an integer spin $S$, corresponding to crossing of an odd 
number of bands, we find that the conductivity vanishes at the crossing, as does the Berry phase \cite{Dora:2011}, while both are finite for half integer spins. 
In three dimensions,  a two band crossing is denoted a Weyl semimetal and is characterized by a topological Chern number instead of a Berry phase.
In that case, the conductance at the nodal point remains finite and scales as
$G_{\textrm{3d}} = \ln 2 \ e^2/(2\pi h) (W/L)^2$ instead of
$G_{\textrm{2d}} =  e^2/(\pi h) (W/L)$ in two dimensions \cite{Baireuther:2014}. Similarly the value of the Fano factor is slightly modified from
$F_{\textrm{2d}} = 1/3$ to  $F_{\textrm{3d}} \simeq 0.57$.
The existence of a finite minimal conductance scaling as $ (W/L)^2$ appears to be a robust property associated with the topological two band crossing as
 its existence is neither modified by a weak disorder \cite{Sbierski:2014}, nor by an anisotropic deformation of the cone or a tilt of this cone which breaks the local
 chiral symmetry \cite{Trescher:2015}.

\section*{Methods}

{\bf Topological characterization.}
The topological characterization of a band crossing point is done by calculating the Berry phase associated to each eigenvector $|\Psi_n \rangle$ upon winding
anticlockwise around the crossing point:
\begin{equation}
 \gamma_n (\vec{K}) = \frac{-i}{\pi} \oint d\vec{q}. \langle \Psi_n | \vec{\nabla}_{\vec{q}} | \Psi_n \rangle  \  .
\label{eq:gamman}
\end{equation}

{\bf Conductance and Fano factor}.
The conductance of a narrow sample is conveniently calculated from the  set of the transmission probabilities $T_n$ of the
conduction channel labeled by $n$ through the Landauer formula
$G =\frac{e^2}{h} \sum_n T_n \ .$
The longitudinal conductivity $\sigma$ is related to this conductance as $ \sigma = LW^{-1} G$.
The Fano factor is related to these transmission coefficients as
$F = \sum_{n} T_{n}(1-T_{n})/\sum_{n} T_{n}$.
The explicit calculation of the transmission probabilities $T_n$
requires solving the Schrödinger equation piecewise and matching the solutions at the boundaries of the sample.

 Numerical calculations of transport were performed using the Kwant code \cite{Groth:2014}, based on a  Green function recursive technique to evaluate the transmission amplitude across a sample.
Typical samples of dimensions $L=100$, $W=300$ in lattice units were used, using a large potential in the electrodes 
$e V_\infty W/(v_F \hbar) = 45$, with Fermi velocity $v_F=3at/(2\hbar)$.

{\bf Acknowledgments}. 
We thank F. Pi{\'e}chon for stimulating discussions about the T$_{3}$ model,  as well as 
A. Akhmerov and X. Waintal for helpful advice with the Kwant package. 
%
This work was supported by the grants ANR Blanc-2010 IsoTop and ANR Blanc-2012 SemiTopo from the french Agence Nationale de la Recherche.


%

\end{document}